\documentclass[floatfix,showpacs,amsmath,amssymb,letterpaper,groupaddresses,superscriptaddress]{article}
\usepackage{graphicx}
\usepackage{latexsym}
\usepackage{amsmath,amssymb}

\usepackage{epsfig}
\usepackage{graphicx}

\usepackage{graphicx}
\usepackage{amssymb,amsmath,times}

\usepackage{color}

\def\be{\begin{equation}}
\def\ee{\end{equation}}
\def\ba{\begin{eqnarray}}
\def\ea{\end{eqnarray}}
\def\la{\langle}
\def\ra{\rangle}

\def\h{\hskip 1cm}

\def\lo{\longrightarrow}

\usepackage{epsfig}
\usepackage{graphicx}

\begin{document}

\begin{titlepage}
\vspace{4cm}
\begin{center}{\Large \bf The Kitaev-Ising model, \\ \vspace{0.5 cm}  Transition between topological and ferromagnetic order}\\
\vspace{2cm} Vahid Karimipour \footnote{Corresponding
author:vahid@sharif.edu},\h Laleh Memarzadeh ,\h Parisa Zarkeshian \\
\vspace{1cm} Department of Physics, Sharif University of
Technology,
\\P.O. Box 11155-9161, Tehran, Iran
\end{center}
\vskip 2cm

%%%%%%%%%%%%%%%%%%%%%%%%%%%%%%%%%%%%%%%%%%%%%%%%%%%%%%

%%%%%%%%%%%%%%%%%%%%%%%%%%%%%%%%%%%%%%%%%%%%%%%%%%%%%%
\begin{abstract}
We study the Kitaev-Ising model, where ferromagnetic Ising
interactions are added to the Kitaev model on a lattice. This
model has two phases which are characterized by topological and
ferromagnetic order. Transitions between these two kinds of order
are then studied on a quasi-one dimensional system, a ladder, and
on a two dimensional periodic lattice, a torus. By exactly
mapping the quasi-one dimensional case to an anisotropic XY chain
we show that the transition occurs at zero $\lambda$ where
$\lambda$ is the strength of the ferromagnetic coupling. In the
two dimensional case the model is mapped to a 2D Ising model in
transverse field, where it shows a transition at finite value of
$\lambda$. A mean field treatment reveals the qualitative
character of the transition and an approximate value for the
transition point. Furthermore with perturbative calculation, we
show that expectation value of Wilson loops behave as expected in
the topological and ferromagnetic phases.
\end{abstract}

\hspace{.3in}
\end{titlepage}

\section{Introduction}
In the study of many body systems, specially those inspired by
quantum computation, a new paradigm is emerging, which embodies
concepts such as topological order, topological phase, and
topological phase transition. Contrary to the traditional Landau
paradigm, topological phases are not characterized by local order
parameters and topological phase transitions are not accompanied
by spontaneous symmetry breaking. In addition to their interest
in condensed matter physics\cite{ref1}, i.e. in fractional
quantum Hall liquids \cite{ref2} and quantum spin liquids
\cite{ref3}, lattice models exhibiting topological order are of
immense interest in the field of quantum computation and
information \cite{ref4}, \cite{ref5} due to their robustness
against decoherence. The simplest such lattice models is the
Kitaev model \cite{ref4}, although other models like color codes
have also been introduced and extensively studied \cite{ref6,
ref6A}. The ground state of the Kitaev model on a surface of genus
$g$, exhibits a $4g$-fold degeneracy which is directly related to
the topology of the surface. Different ground states look exactly
the same if probed by expectation values of local observable and
are only distinguished if probed by global string-like operators
going around non-trivial homology cycles of the surface. One can
thus use these ground states to encode $2g$ qubits which are
robust against errors and decoherence. Moreover one can do
topological computations on these qubit states if one uses
braiding and
fusion of anyonic excitations of these models. \\

It is then natural to ask how much this topological order in the
original Kitaev model or its generalizations to $Z_N$ group or
the topological color codes \cite{ref6, ref6A} are resilient
against various kinds of perturbations \cite{ref7, Hamma1, Hamma2,
Jahromi, Shulz, Dusuel}, temperature fluctuations \cite{ref8} and
so on. For example one can imagine that a strong magnetic field
will eventually align all the spins in the direction of the
magnetic field and the topological ordered phase transforms to a
spin-polarized phase, \cite{Hamma2, Vidal1, Vidal2} a phase which
is easily recognized by local measurements of spins. Or one can
imagine that at high enough temperature the topological phase
transforms to a disordered phase \cite{ref10}, again recognizable
locally. In these transitions a topologically degenerate ground
space transforms to other forms of ground states. Phase
transitions of this
kind have been studied in \cite{ref7, Hamma1, Hamma2, Jahromi, Shulz, Dusuel, Vidal1, Vidal2, ref8, new1, new2, new3, Pas}. \\

It is the aim of this paper to study another kind of transition
in these models.  For concreteness we take the Kitaev model and
ask how topological order can transform to ferromagnetic order.
This transition is induced by Ising interaction and moreover it
signifies a transition between two kinds of degenerate ground
states. That is in the Kitaev limit the degeneracy comes from
topology and in the Ising limit, it comes from symmetry. This
will certainly add to our knowledge about topological order and
the way it is either destroyed (i.e. by temperature or by
magnetic fields) or changes to other types of local order (i.e.
by ferromagnetic interactions).  \\

To this end, we introduce a model in which Ising terms compete
with Kitaev interactions, one to establish ferromagnetic order
and the other to establish topological order. We study the model
on both 2 dimensional torus and the quasi-one dimensional ladder
network \cite{ref12} which has almost all the characteristics of a
topological model, i.e. topological degeneracy, robustness and
anyonic excitations. Studying two different models has the
benefit of understanding the role of dimension in this
transition. To find the possibility of the transition and the
transition point if any, we exactly map the problem of finding
the ground state of the model to a simpler problem. In the ladder
case, we map the model to a one-dimensional XY model, whose
anisotropy is tuned by the ration of Ising to Kitaev couplings.
This model is exactly solvable by free fermion techniques, its
ground state is non-degenerate and smoothly varying, except at
the extreme points (XX or YY interaction). Therefore in the case
of ladder, there is no transition at finite Ising coupling.
However in the 2D case, we exactly map the problem to the 2D
Ising model in transverse field. The latter model has been
studied using different methods \cite{ref13, ref13new,
ref13newnew} and is known to show a quantum phase transition. We
show that the two sides of transition point correspond to
topological and ferromagnetic order in the Kitaev-Ising model.
This provides strong evidence for a transition between these two
phases in the
original model.\\

The structure of the paper is as follows: In section \ref{sec2}
we briefly review some preliminary facts on Kitaev model,
emphasizing their difference on the torus and on the ladder. In
section \ref{sec3} we introduce the Kitaev-Ising model and solve
it exactly on the ladder in subsection \ref{subsec41}. In
subsection \ref{subsec42}, we map the Kitaev-Ising model to 2D
Ising model in transverse field and analyze the degeneracy
structure of the model and interpret it in terms of the original
model. In an appendix, by a simple mean field analysis, we find
the transition point which turns out to be near the actual one
obtained by more accurate numerical means \cite{ref14, ref13,
ref13new, ref13newnew}. To substantiate the idea of a phase
transition between topological and non-topological phases, in
section {\ref{wilson}} we use the above mapping which facilitates
an estimation of the expectation values of Wilson loops in the two
regimes. These estimates indeed turns out to be as we expect,
that is, the expectation value of a Wilson loop $\la W_C\ra$
behaves as the exponential of a quantity which is proportional to
the perimeter of the $C$ near the Kitaev point and to the area
enclosed by $C$ near the Ising point. The paper concludes with a
discussion.

\section{A brief account of the Kitaev Model}\label{sec2}
In this section we briefly review the Kitaev model \cite{ref4} in
order to set up the notation and use its main concepts in the
sequel. Consider a lattice whose set of vertices, edges and
plaquettes are respectively denoted by $V$, $E$ and $P$
respectively. The number of elements in these sets are
respectively denoted by $|V|, |E|$ and $|P|$ respectively. Spin
one-half particles live on the edges of this lattice and hence
the dimension of the full Hilbert space is given by $2^{|E|}$. The
Kitaev Hamiltonian on this lattice is given by

\begin{equation}\label{1}
  H_{Kitaev}:=-J\sum_{s\in V} A_s - K \sum_{p\in P} B_p
\end{equation}
where
\begin{equation}\label{2}
A_s:=\prod_{i\in s}\sigma_{i,x}, \h B_p:=\prod_{i\in \partial
p}\sigma_{i,z}.
\end{equation}
Here $i\in s$ means the edges incident on a vertex $s$ and $i\in
\partial p$ means the edges on the boundary of a plaquette $p$.
The coupling constants, $J, K$ and are taken to be positive. It
is easily verified that all the vertex and plaquette operators
commute with each other and square to the identity operator. The
ground state is thus the common eigenvectors of all the vertex
and plaquette operators with eigenvalue 1, that is $|\Phi\ra$ is
a ground state of the Kitaev model if it satisfies $
  A_s|\Phi\ra=|\Phi\ra, \ \ \ B_p|\Phi\ra=|\Phi\ra, \ \ \  \forall \ s,
  p.
$

\subsection{Kitaev Model on a two-dimensional periodic lattice (a torus)}\label{subsec21}
For a 2D rectangular lattice of $N$ vertices, with periodic
boundary conditions on both directions ( a torus), we have
$|V|=N, |E|=2N, $ and $|P|=N$. Hence the dimension of the Hilbert
space is $2^{2N}$ and we also have $N$ vertex operators $A_s$ and
$N$ plaquette operators $B_p$ all commuting with each other and
with the Hamiltonian. However there are two global constraints on
the torus, namely
\begin{equation}\label{5}
  \prod_s A_s=I, \h \prod_pB_p=I,
\end{equation}
leading to ${2N-2}$ independent commuting operators and hence a
4-fold degeneracy of the ground state. In fact one notes that
there are four string operators all commuting with the
Hamiltonian, which are defined as follows:
\begin{equation}\label{6}
  T_z^{(1)}:=\prod_{i\in C_1}\sigma_{i,z}, \h  T_z^2:=\prod_{i\in C_2}\sigma_{i,z},
\end{equation}
\begin{equation}\label{7}
  T_x^{(1)}:=\prod_{i\in \tilde{C_1}}\sigma_{i,x}, \h  T_x^2:=\prod_{i\in \tilde{C_2}}\sigma_{i,x},
\end{equation}
where $C_1$ and $C_2$ are two homology cycles along the edges of
lattice of the torus and $\tilde{C_1}$ and $\tilde{C_2}$ are two
cycles running around the dual lattice (figure 1). Note that
these operators, corresponding to homology cycles (curves which
do not enclose any area) cannot be expressed in terms of vertex
and plaquette operators. They have the following relations with
each other:
\begin{equation}\label{8}
  T_x^{1}T_z^{1}=-T_z^1T_x^1, \h T_x^2T_z^2=-T_z^2T_x^2.
\end{equation}
while all other relations are commutative ones. In other words,
the operators $(T_x^1,T_z^1)$ and $(T_x^2,T_z^2)$ form two copies
of the Pauli operators $\sigma_x$ and $\sigma_z$ which act to
distinguish the four degenerate ground states of the Kitaev model
and turn them into each other. That is, if we denote the four
ground states by $|\Phi_{s_1,s_2}\ra, s_1,s_2=0,1$, then we
have
\begin{equation}\label{9}
  T_z^{1}|\Phi_{s_1,s_2}\ra=(-1)^{s_1}|\Phi_{s_1,s_2}\ra, \h  T_z^{2}|\Phi_{s_1,s_2}\ra=(-1)^{s_2}|\Phi_{s_1,s_2}\ra,
\end{equation}
and
\begin{equation}\label{10}
  T_x^{1}|\Phi_{s_1,s_2}\ra=|\Phi_{s_1+1,s_2}\ra, \h  T_x^{2}|\Phi_{s_1,s_2}\ra=|\Phi_{s_1,s_2+1}\ra.
\end{equation}

\begin{figure}[t]
\centering\vspace{1cm}
\includegraphics[width=6cm,height=5.3cm,angle=0]{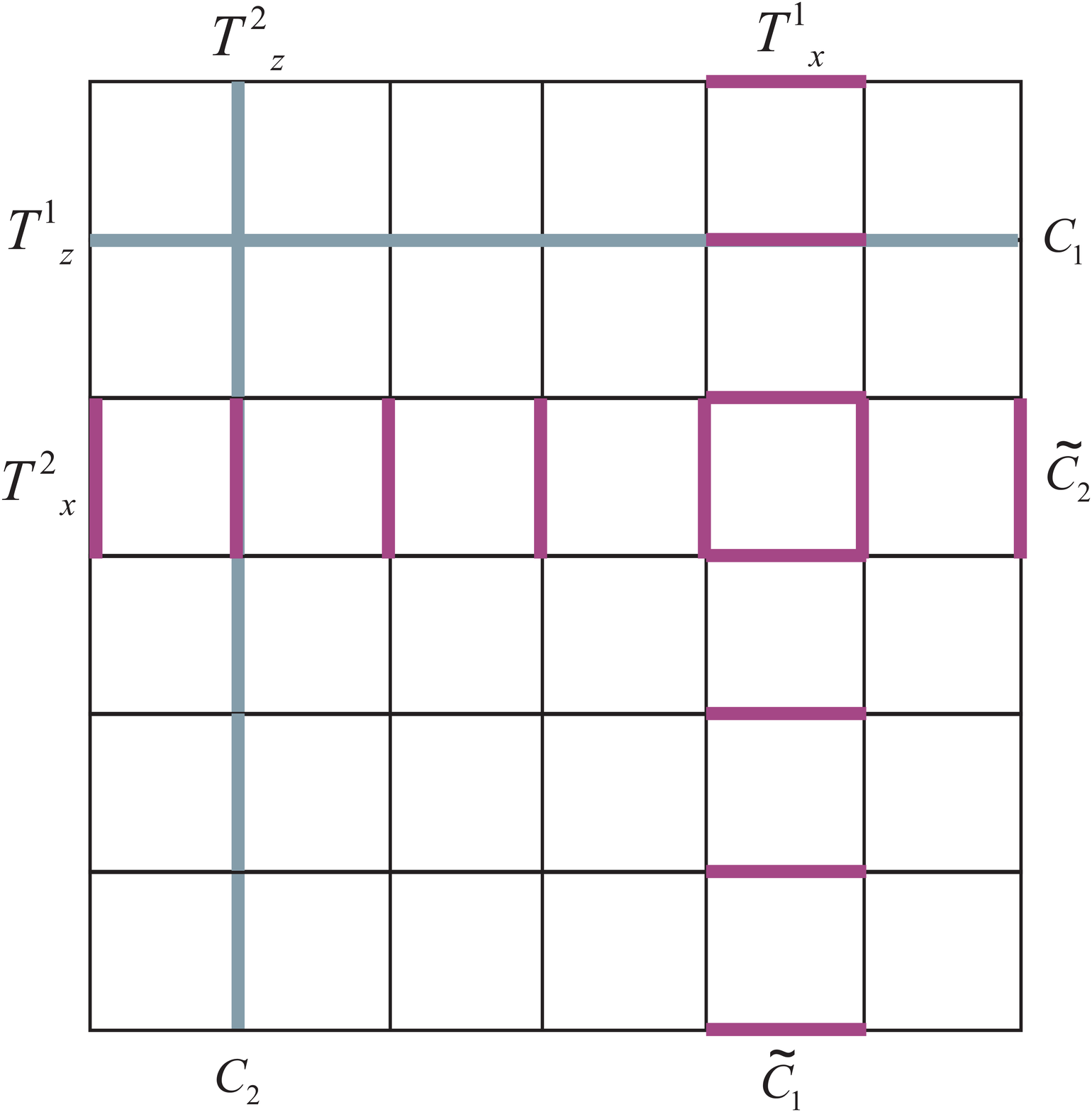}\vspace{1cm}
\caption{(Color Online) The string operators $T_z^1,T_z^2$ and
$T_x^1, T_x^2$. All of them commute with the pure Kitaev
Hamiltonian ($ H(\lambda=0)$). Only  $T_z^1$ and $T_z^2$ commute
with $H(\lambda)$ for all $\lambda$. }
\end{figure}\label{fig1}

\begin{figure}[t]
\centering\vspace{1cm}
\includegraphics[width=7cm,height=7cm,angle=0]{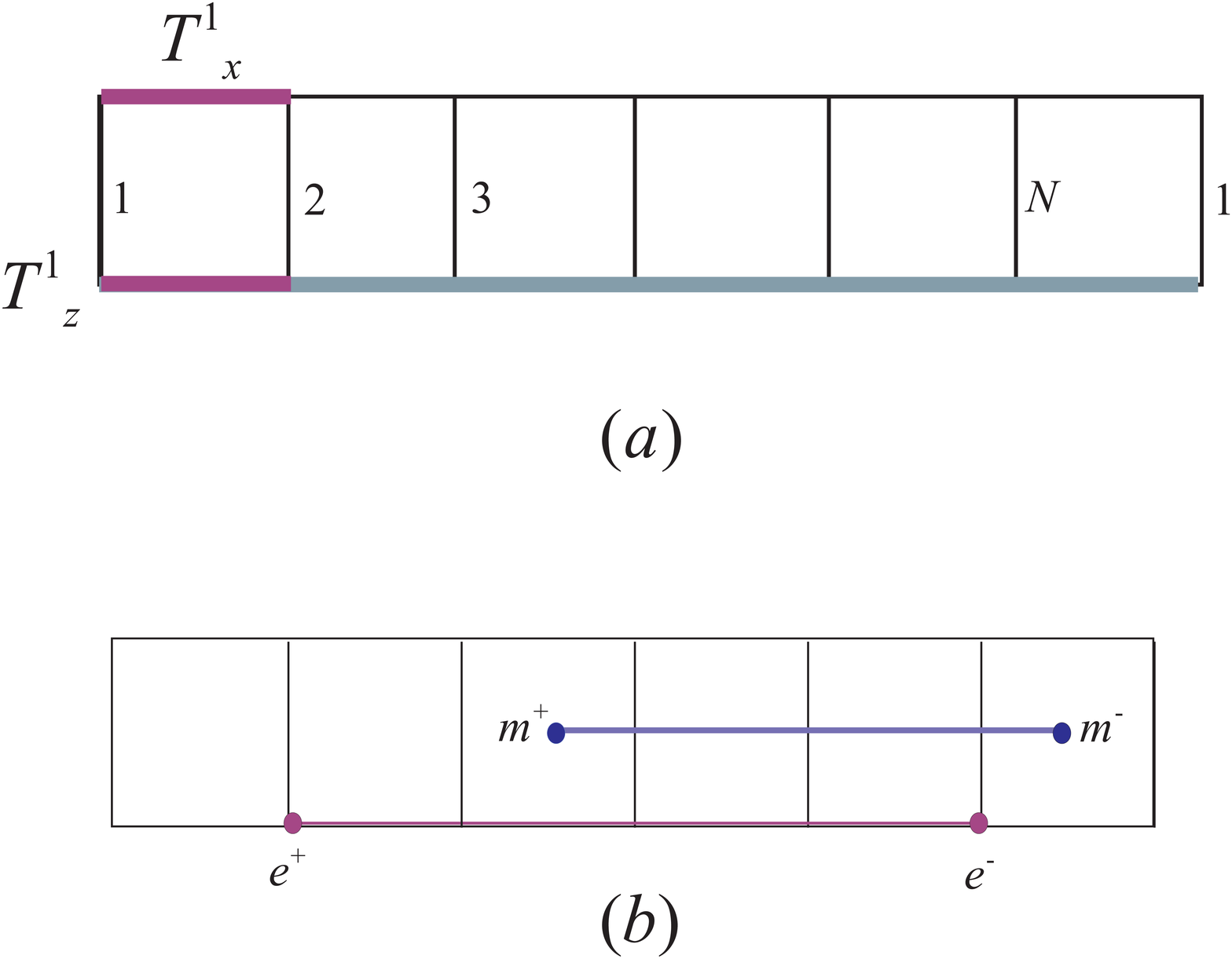}\vspace{1cm}
\caption{(Color Online) (a) The operators $T_z^1$ and $T_x^1$ on
the ladder. $T_z^1$ is the product of $\sigma_{i,z}$ on the lower
leg, while $T_x^1$ is the product of two $\sigma_{x}$'s one on
each leg. Both commute with $H(\lambda=0)$, but only $T_z^1$
commutes with $H(\lambda)$. $T_x^1$ contains edges from the two
legs. No operator containing only edges from one leg can
substitute it. (b) Electric and magnetic anyons on the ladder.}
\end{figure}\label{fig2-LadderSimple}

\subsection{The Kitaev Model on the quasi-one dimensional lattice (a ladder)}\label{subsec22}
Since we will also study the Kitae-Ising Hamiltonian on the
quasi-one dimensional systems, it is in order to note a few minor
differences that the Kitaev model on the ladder has with the 2D
case. Consider a ladder, as shown in figure (2), with $N$
plaquettes.  There are $2N$ vertices and  $3N$ edges. So the
dimension of the Hilbert space is $2^{3 N}$. We have periodic
boundary condition only along the legs. The number of independent
operators $B_p$ is equal to $N$, while the number of independent
$A_s$ operators is $2N-1$, since $\prod_{s}A_s=I$. (There is no
such constraint on the $B_p$ operators on the ladder). Therefore
the total number of independent commuting operators is equal to
$3N-1$ leading to a 2-fold degeneracy for the ground state. On
the ladder only one pair of operators in (\ref{6}, \ref{7})) with
their properties remain, which are denoted by $T_z^1$ and $T_x^1$
in figure (2). In fact the analogue of operator $T_z^2$ is an
operator like $\sigma_{1,z}$ sitting on a single rung of the
ladder, which no longer commutes with the Hamiltonian and the
analog of operator $T_x^2$ is no longer independent from the
vertex operators, since $T_x^2=\sigma_{1,x}\sigma_{2,x}\cdots
\sigma_{N,x}=A_1A_2\cdots A_N$, where $A_i$ denote the vertex
operators on the upper (or lower) leg of the ladder. This is in
accord with the two-fold degeneracy of the ladder, that is if we
denote the two ground states of the ladder by $|\Psi_s\ra, \
s=0,1$, then we have

\begin{equation}\label{11}
  T_z^{1}|\Phi_{s}\ra=(-1)^{s}|\Phi_{s}\ra,\h T_x^{1}|\Phi_{s}\ra=|\Phi_{s+1}\ra.
\end{equation}
Note that the Kitaev model on a ladder, being a quasi-one
dimensional system allows a restricted form of topological order.
That is, concepts like area law for Wilson loops, or topological
entanglement entropy may not apply to it. However there are still
some topological characteristics in the ground states. First we
have ground state degeneracy which does not come from symmetry.
The two states being converted to each other by the global string
operator $T^1_x$. Second we have a finite gap. Also the
expectation value of any local operator, i.e one which does NOT
traverse the the two legs of the ladder, is the same on the two
ground states $|\Phi_0\ra$ and $|\Phi_1\ra$. In fact an operator
which distinguishes the two ground states, should be one which
commutes with the Hamiltonian and at the same time anti-commutes
with $T^1_z$. Such an operator is given by $T_x^1$ which
necessarily contains both legs of the ladder. (It is in this
sense that no local operator (one defined on a single leg) can
distinguishes the two ground states).  Finally the system has
anyonic exitations of electric and magnetic charges with integer
charges and abelian statistics. In fact as shown in figure (2),
an open string of $\sigma_z$ operators along the edges creates
electric anyons at the end points, while an open string of
$\sigma_x$ operators along the rungs creates magnetic anyons and
cycling any electric anyon around a magnetic one creates a phase
of (-1). Again the specific topology of the ladder reflects
itself in the properties of its anyons in that, only electric
anyons can move around the magnetic anyons. \\

Therefore many of the concepts pertaining to topological order
are valid also for this quasi-one dimensional system. When we
speak of topological order on the ladder, we mean this restricted
meaning of the word. On 2D we do not have such a restriction. \\

\section{The Kitaev-Ising model }\label{sec3}
We define the Kitaev-Ising Hamiltonian on any lattice as follows
\begin{equation}\label{13}
  H(\lambda):=H_{Kitaev}+\lambda H_{Ising},
\end{equation}

in which $H_{Kitaev}$ is the usual Kitaev Hamiltonian \eqref{1} and
$H_{Ising}$ is the Ising interaction between nearest neighbor
links
\begin{equation}\label{14}
   H_{Ising}= -\sum_{\la i,j\ra} \sigma_{i,z}\sigma_{j,z},
\end{equation}
where $\la i,j\ra$ means nearest-neighbor edges on the lattice.
It is important to note that in the presence of the Ising
interaction, the plaquette operators still commute with the full
Hamiltonian, although the vertex operators no more do so:
\begin{equation}\label{15}
  [B_p, H(\lambda)]=0, \ \ \  \forall p, \h [A_s, H(\lambda)]\ne
  0.
\end{equation}
Moreover from the four string operators, shown in figure (1), which
commutes with the Kitaev Hamiltonian, only two retain this
property in the presence of Ising interaction, namely

\begin{equation}\label{16}
  [T_z^1,H(\lambda)]=0 \h [T_z^2,
  H(\lambda)]=0,
\end{equation}
but
\begin{equation}\label{17}
  [T_x^1,H(\lambda)]\ne 0 \h [T_x^2,
  H(\lambda)]\ne 0.
\end{equation}

Correspondingly for the ladder, only the operator $T_z^1$ is
defined which commutes with the Hamiltonian. \\

In view of the fact that $[B_p,H(\lambda)]=0, \ \ \forall p, $
and $ \lambda $ and the fact that in both limits (pure Kitaev and
pure Ising) the ground states have eigenvalue $+1$ for all
$B_p$'s, we conclude that the ground states of the Kitaev-Ising
model lie in the subspace where $B_p=1$ for all the plaquettes.
Denoting this subspace by ${\cal V}_0$,
\begin{equation}\label{18}
  {\cal V}_0:=\{|\phi\ra \ , \  B_p|\phi\ra=|\phi\ra\}.
\end{equation}
Therefore the restriction of the Hamiltonian to this subspace,
$H_0(\lambda):=H(\lambda)\mid_{{\cal V}_0}$ is given by
\begin{equation}\label{19}
  H_0(\lambda)=-J\sum_s A_s-\lambda \sum_{\la i,j\ra}
\sigma_{z,i}\sigma_{z,j}-K|P|,
\end{equation}
where $|P|$ is the number of plaquettes in the lattice. Therefore
the Ising coupling $\lambda$ or more precisely the ration
$\frac{\lambda}{J}$ tunes the competition of ferromagnetic order
and topological order. When this ration is zero we have pure
Kitaev model and topological order, and when it is very strong,
we have ferromagnetic order. In both cases we have degeneracy,
but in one case the degeneracy is due to topology and in the
other case it is due to symmetry. It is also interesting to note
that the degeneracy of the ferromagnetic order is always
two-fold, i.e. either all the spins are up or all are down, while
the topological degeneracy is four-fold for the torus and
two-fold for the ladder. In the subsequent sections we will
understand how this order and the corresponding degeneracy
changes as we change the parameter $\frac{\lambda}{J}$.

\section{ Solution of the Kitaev-Ising model}\label{sec4}
We showed that the ground states of the Kitaev-Ising model live in
the subspace ${\cal V}_0$ defined in \eqref{18}. The restriction
of $H(\lambda)$ to this subspace is given by \eqref{19}. To
further diagonalize $H_0(\lambda)$, we construct a suitable basis
for the subspace ${\cal V}_0$ and through this we transform
$H_0(\lambda)$ to very simple models which have been studied
previously. In fact, we will show that for the ladder,
$H_0(\lambda)$ turns out to be the Hamiltonian of a
one-dimensional XY chain, while for the 2D lattice,
$H_0(\lambda)$ is the Hamiltonian of an Ising model in transverse
field. \\

The way this basis is constructed is of utmost importance, in
fact it should be constructed in such a way that all the
operators in the Hamiltonian, i.e. the vertex and plaquette  and
also the Ising terms should be represented by nearest neighbor
interactions between Pauli operators on virtual spins. Otherwise,
one may come up with an inappropriate reduced Hamiltonian, one
which may entail three or four-spin interactions or in case of
two-body interactions it may entail longer than nearest-neighbor
interactions. In other words, choosing this basis is a
significant step in the process of diagonalization. Due to the
difference between the topology of the ladder and the torus, we
proceed in two different ways in construction of this basis. We
start with
the ladder and then study the case of 2D torus. \\

\subsection{On the ladder}\label{subsec41}
Consider the ladder shown in figure (3), where we take for
definiteness the number of plaquettes to be an even number. We
first note that the state
\begin{equation}\label{20}
  |\Phi_0\ra:=\prod_{i=1}^N(1+B_i)|+\ra^{\otimes 3N},
\end{equation}
where $|+\ra$ is the positive eigenstate of $\sigma_x$, is a
ground state of the pure Kitaev model, (one can easily check that
it satisfies $A_s|\Phi\ra=B_p|\Phi\ra=|\phi\ra$ for all $s$ and
$p$). Consider the curve $C_1'$ on the ladder (shown in figure
(3)). This is a cycle going around the ladder and in fact it is
equivalent to the straight curve $C_1$ shown in figure (2) (this
equivalence is explained below). Therefore the other ground state
of the pure Kitaev model on the ladder is nothing but
\begin{equation}\label{21}
  |\Phi_1\ra:=\prod_{i\in C_1'}\sigma_{i,z}|\Phi_0\ra.
\end{equation}
By equivalence of $C'_1$ and $C_1$ we mean that the difference of
$\prod_{i\in C'_1}\sigma_{i,z}$ and $T_{z}^1:=\prod_{i\in
C_1}\sigma_{i,z}$ is a product of $B_i$ operators which has no
effect on $|\Phi_0\ra$. In fact we can simply straighten a $
\sqcap$ or a $\sqcup$ by multiplying with the $B$ inside them. We
now construct the following set of un-normalized states
\begin{equation}\label{22}
 |\tilde{{\bf r}}\ra:= |\tilde{r}_1, \tilde{r}_2, \tilde{r}_3, \cdots \tilde{r}_{2 N}\ra=\prod_{i\in C'_1}^{2{N}}\sigma_{i,z}^{r_i}|\Phi_0\ra, \ \  \ \
 r_i=0,1.
\end{equation}

Clearly these states satisfy $B_p|\tilde{{\bf r}}\ra=|\tilde{{\bf
r}}\ra$ for all $p$. We also note that
$$|\{\tilde{r}_i=0\}\ra=|\Phi_0\ra \ \ \ , \ \ \
|\{\tilde{r}_i=1\}\ra=|\Phi_1\ra.
$$ Moreover, they are orthogonal. For the proof of
orthogonality, the basic idea to use, is that $|\Phi_0\ra$ can be
viewed simply as a linear combination of closed loops of spin $-$
particles in a background of all spin $+$ particles. Now if ${\bf
r}\ne {\bf r'}$, then it is easy to see that $\la \tilde{{\bf
r}}|\tilde{{\bf r}}'\ra\equiv \la \Phi_0|\prod_{i\in
C'_1}\sigma_{i,z}^{r_i-r'_i}|\Phi_0\ra$ is the product of two
states where one ($|\Phi_0\ra$) has all closed loops and the
other has open strings of spin down states, the product of which
is zero. Finally their number is $2^{2N}$ which is equal to
dimension of ${\cal V}_0$. Thus with proper normalization, they
form an orthonormal basis for ${\cal V}_0$.
\begin{figure}[t]
\centering\vspace{1cm}
\includegraphics[width=10cm,height=2cm,angle=0]{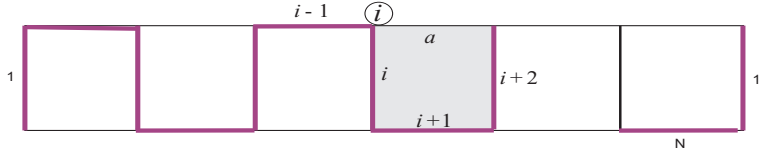}\vspace{1cm}
\caption{(Color Online) The curve used for generating an
orthonormal basis in the sector ${\cal V}_0$ for the ladder
network. }
\end{figure}\label{fig3-SubLadderFig}
Why we have chosen this particular curve and this particular form
for expressing the states of this sector? The answer lies in the
nice form (i.e. nearest-neighbor two-body interaction) of the
reduced Hamiltonian $H_0(\lambda)$. If we choose the curve as a
simple straight form like $C_1$, then the states $|{\bf r}\ra$ do
not span the whole subspace ${\cal V}_0$. \\

To find $H_0(\lambda)$ in this new basis, we should determine the
action of operators $A_s$ and also the Ising terms on these basis
states. Due to the zigzag shape of the path and the appearance of
the corresponding operators in the definition of $|\tilde{{\bf
r}}\ra$, we find that any vertex operator like $A_i$ in figure
(3) when acting on the state \eqref{22} passes through all the
operators except $\sigma_{z,i-1}$ and $\sigma_{z,i}$ with which
it anticommutes. Thus the passage of $A_i$ through the whole
chain of operators produces only the factor $(-1)^{r_{i-1}+r_i}$,
hence the following effective operation on the basis states:

\begin{equation}\label{23}
  A_i|\tilde{{\bf r}}\ra=Z_{i-1}Z_i|\tilde{{\bf r}}\ra,
\end{equation}
where $Z_i$ is the notation of Pauli operator $\sigma_z$ in this
subspace.\\

\textbf{Notation:} Original Qubit states on the edges of the
lattice are denoted without a $\tilde{}$. Thus $|0\ra$ and
$|1\ra$ denote the computational basis states on the edges,
$|0\ra=|z,+\ra$ and $|1\ra=|z,-\ra$. Pauli operators on these
qubits are denoted by $\sigma_x$ and $\sigma_z$. The qubit states
in \eqref{22} are always denoted by a $\tilde{}$ and the corresponding
Pauli operators on these qubits are denoted by capital letters,
$X$ and $Z$.\\

We now come to the Ising terms. Consider a group of Ising terms
in one plaquette, the one shaded in figure (3). This can be
written as
\begin{equation}\label{24}
C_i:=\sigma_{i,z}\sigma_{i+1,z}+\sigma_{i,z}\sigma_{a,z}+\sigma_{i+1,z}\sigma_{i+2,z}+\sigma_{a,z}\sigma_{i+2,z}.
\end{equation}
To express the action of $C_i$ in the basis (\ref{22}), we note
that this can be rewritten as
\begin{equation}\label{25}
C_i:=\left(\sigma_{i,z}\sigma_{i+1,z}+\sigma_{i+1,z}\sigma_{i+2,z}\right)(1+B_i),
\end{equation}
where $B_i$ is the operator corresponding to the same (shaded)
plaquette. The operator $(1+B_i)$ gives a factor of $2$ when
acting on the state $|\tilde{{\bf r}}\ra$ and the remaining
$\sigma_{z,i}$ operators only flip the corresponding bit labels
$\tilde{r}_i$. Hence the following effective action on the state:

\begin{equation}\label{26}
  C_i|\tilde{{\bf r}}\ra=2\left(X_{i}X_{i+1}+X_{i+1}X_{i+2}\right)|\tilde{{\bf
  r}}\ra,
\end{equation}
where again $X_i$ is used to denote the first Pauli operator on
the subspace ${\cal V}_0$. Putting everything together, we arrive
at the following effective Hamiltonian on this subspace:
\begin{equation}\label{27}
  H_0(\lambda)=-J\sum_{i}Z_iZ_{i+1}-2\lambda\sum_i X_{i}X_{i+1}-K{N}.
\end{equation}

This is an XY Hamiltonian in the absence of external magnetic
field which has been studied extensively in the literature \cite{ref15}.
 Its exact solution is provided by turning it into a free
fermion model by Jordan-Wigner and Bogoluibov transformations.
Its ground state is non-degenerate except at the extreme points
$\lambda=0$ or $\lambda\lo\infty$. The ground state and the
correlation functions show no non-analytical behaviour and no
quantum phase transition occurs for finite $\lambda$. The only
thing which happens is that the two-fold degeneracy breaks for
any value of $\lambda$ except at the extreme points $\lambda=0,$
(Pure Kitaev) or $ \lambda\lo \infty$ (Pure Ising). The end
conclusion is that a transition from topological to ferromagnetic
order does not occur for finite $\lambda$ in quasi-one
dimensional systems.  \\

Before leaving the subject of ladders, it is instructive to have
a final look at the ground states of $H_0(\lambda)$ at the two
extreme points. In these two limits, the ground state(s) of
\eqref{27} should have simple product form (in terms of the labels
$\tilde{r}_i$). Let us see if these are really what we expect for
the Kitaev-Ising model.  In the limit $\lambda=0$, equation
\eqref{27} says that the virtual spins should all align either in
the positive or negative $z$ direction, hence there are two
degenerate ground states given by $|\tilde{\bf{0}}\ra$ and
$|\tilde{\bf{1}}\ra$ respectively. As explained at the beginning
of this subsection, these two states are clearly the two Kitaev
states $|\Phi_0\ra$ and $|\Phi_1\ra$ on the ladder. The other
limit, however, is more tricky to show. In the limit $\lambda\lo
\infty$ all the virtual spins should align either in the positive
$x$ or negative $x$ direction. We should show that this means
that the actual spins on the edges of the ladder all align in the
positive or negative $z$ direction. To see this consider the
state $|\tilde{+},\tilde{+},\cdots \tilde{+}\ra$. In view of the
definition \eqref{22} and the structure of \eqref{20}, and the
fact that $|\tilde{+}\ra\propto |\tilde{0}\ra+|\tilde{1}\ra$,
this corresponds to

\begin{equation}\label{28}
  |\tilde{+},\tilde{+},\cdots \tilde{+}\ra=\prod_{i\in C'_1} (1+\sigma_{i,z})|\Phi_0\ra =\prod_{i=1}^{N} (1+B_i)
  |\Omega_0\ra,
\end{equation}
where $|\Omega_0\ra:=\prod_{i\in
C'_1}(1+\sigma_{i,z})|+\ra^{\otimes 3N}$ is a state on the ladder
which we depict in figure (4). Here we have used the property
$(1+\sigma_z)|+\ra=|0\ra$.  When the operators $1+B_i$ act on
$|\Omega_0\ra$, they
turn the remaining $+$ states into $0$ and hence turn it into a ground state of the pure Ising model.\\

\begin{figure}[t]
\centering\vspace{1cm}
\includegraphics[width=11cm,height=3cm,angle=0]{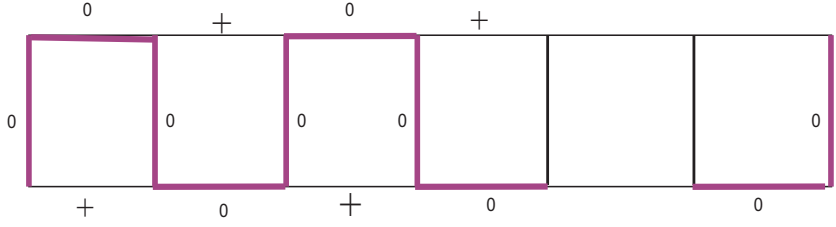}\vspace{1cm}
\caption{(Color Online) The state $\Omega_0$, defined in equation
\eqref{28} and the following paragraph.}
\end{figure}\label{fig4-Omega0fig}

For the other state $|\tilde{-}\ra$ a similar reasoning works in
view of $(1-\sigma_{z})|+\ra=|1\ra$ where $|1\ra$ is
spin down in the z-direction.\\

\subsection{On the two dimensional lattice}\label{subsec42}
%limit-degenerate-sublattices

We now turn to the square lattice  and show that $H_0(\lambda)$,
describing the interactions of virtual spins, is in fact the
Hamiltonian of a  2D Ising model in transverse magnetic field.
This model is known to undergo a transition from ferromagnetic
order to spin-polarized ordered phase. These phases, are shown to
correspond
respectively to topological and ferromagnetic ordered phases for the actual spins on the edges of the lattice. \\

To show this equivalence, we follow steps similar to the ones in
previous section, however to represent the Hamiltonian in a
simple form, we should choose an entirely different basis for the
subspace ${\cal V}_0$.
\begin{figure}[t]
\centering\vspace{1cm}
\includegraphics[width=6cm,height=5.3cm,angle=0]{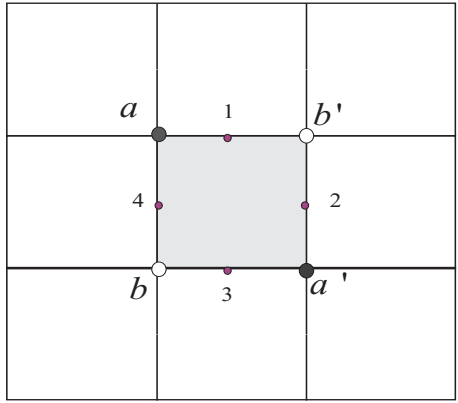}\vspace{1cm}
\caption{(Color Online) A portion of the lattice with their
Kitaev and Ising interactions.  The Ising interactions are
between nearest-neighbor links. An Ising interaction like
$\sigma_{3,z}\sigma_{4,z}$ (shown with the dash line) commutes
with all the vertex and plaquette operators except with $A_{a}$
and $A_{a'}$, to which it anti-commutes. Similarly an Ising term
like $\sigma_{2,z}\sigma_{3,z}$ commutes with all but $A_b$ and
$A_{b'}$.}
\end{figure}\label{fig5-new2DOblique}
Let the lattice have $N$ plaquettes. Then the number of edges
will be $2N$ and the number of vertices will be $N$. In a concise
notation we have $|P|=N$, $|E|=2N$, and $|V|=N$. The dimension of
the full Hilbert space is thus $2^{2N}$. ${\cal V}_0$ is the
common eigenspace of all $B_p$ operators with eigenvalue $+1$.
Since the number of independent plaquette operators is $N-1$,
this means that $dim({\cal V}_0)=\frac{2^{2N}}{2^{N-1}}=2^{N+1}$.
Furthermore this subspace is decomposed to four different
disconnected subspaces according to the eigenvalues of the global
string operators $T_z^{1}$ and $T_z^2$. Let us denote this
decomposition by

\begin{equation}\label{29}
  {\cal V}_0={\cal V}_0^{++}\oplus {\cal V}_0^{+-}\oplus {\cal V}_0^{-+}\oplus{\cal
  V}_0^{--}.
\end{equation}

Each subspace is $2^{N-1}$ dimensional. Consider now the states
\begin{equation}\label{30}
  |\tilde{{\bf r}}\ra:=\prod_{i\in V} A_i^{r_i}|0\ra^{\otimes |E|}, \h
  r_i=0,1.
\end{equation}
Obviously these states satisfy $B_p|\tilde{{\bf
r}}\ra=|\tilde{{\bf r}}\ra$. Moreover, when an $A_i$ acts on these
states, it increases (by mod 2) the label $r_i$, hence the action
of each $A_i$ on these states is represented by the bit-flip Palui
operator $X_i$. In view of the constraint $\prod_{i\in E}A_i=1$,
we have the equality $|\tilde{{\bf r}}\ra=|\overline{\tilde{\bf
r}}\ra$, where $\overline{r_i}=r_i+1, \\  mod\  2 \ \forall i$.
The subspace ${\cal V}_0^{++}$ is therefore span of the
equivalence class of states $[|\tilde{{\bf r}}\ra]=(|\tilde{{\bf
r}}\ra,|\overline{\tilde{{\bf r}}}\ra)$. The other subspaces can
be constructed similarly, i.e. ${\cal V}_0^{-+}=span
\{[T_x^1|\tilde{{\bf r}}
\ra]=(T_x^1|\tilde{{\bf r}}\ra,T_x^1|\overline{\tilde{{\bf r}}}\ra)\}$.\\

When $\lambda\lo \infty$, (pure Ising model), the two ground
states of the pure Ising model are clearly in the subspace ${\cal
V}_0^{++}$, hence by the fact that $[T_z^{1}, H(\lambda)]=[T_z^2,
H(\lambda)]=0$ and by continuity we find that the ground states
of $H(\lambda)$ live in
the subspace ${\cal V}_0^{++}$. Hereafter we will focus on this subspace.\\

To proceed we also assume that the lattice is bi-partite, i.e.
$V=V_A\cup V_B$, where the vertices in $V_A$ are denoted by black
circles in figure (5) and those of $V_B$ are denoted by white
circles. Note that this puts a condition of even number of
vertices in both directions.  We need to find the action of the
operators $A_i$ and the Ising terms on the states (\ref{30}).
 It is
obvious that the action of a vertex operator like $A_i$ on the
state \eqref{30} is to simply flip the bit $r_i$, therefore $A_i$
acts on this subspace as $X_i$. Next we come to the Ising
interactions. Consider the shaded plaquette in figure (5). The
Ising interactions are given by
\begin{equation}\label{31}
  \sigma_{1,z}\sigma_{2,z}+\sigma_{2,z}\sigma_{3,z}+\sigma_{3,z}\sigma_{4,z}+\sigma_{4,z}\sigma_{1,z}=
  (\sigma_{1,z}\sigma_{2,z}+\sigma_{1,z}\sigma_{4,z})(1+B),
\end{equation}
where $B$ is the plaquette operator containing the links 1, 2, 3,
and 4. We now use the fact that an Ising interaction like
$\sigma_{1,z}\sigma_{2,z}$ commutes with all the vertex operators
and anit-commute with $A_a$ and $A_{a'}$. This means that when
$\sigma_{1,z}\sigma_{2,z}$ acts on the state $|\tilde{{\bf
r}}\ra$ it simply produces a factor $(-1)^{r_a+r_{a'}}$, that is
this operator acts on the subspace ${\cal V}_0^{++}$ as
$Z_aZ_{a'}$. Similarly the Ising term $\sigma_{1,z}\sigma_{4,z}$
commutes with all the vertex operators and anti-commutes with
$A_b$ and $A_{b'}$ and with the same reasoning the action of this
operator on ${\cal V}_0^{++}$ is equivalent to $Z_bZ_{b'}$.
Therefore the Ising terms couple the nearest neighbor vertices of
the sublattice $V_A$ and $V_B$ separately. Putting everything
together we find the following effective Hamiltonian:
\begin{equation}\label{32}
  H=H_A+H_B-KN,
\end{equation}
where $KN$ comes from the action of $\sum_p B_p$ on ${\cal V}_0$
and $H_A$ and $H_B$ are each a 2D Ising model in transverse field
on sublattice $A$ and $B$ respectively:
\begin{equation}\label{33}
  H_A=-J\sum_{i\in A} X_i - 2\lambda \sum_{\la i,j\ra\in A} Z_iZ_j
\end{equation}
and
\begin{equation}\label{34}
  H_B=-J\sum_{i\in B} X_i - 2\lambda \sum_{\la i,j\ra\in B}
  Z_iZ_j.
\end{equation}

Here $\la i,j\ra$ means nearest-neighbor vertices on the
corresponding sublattice.  Note that the factor of $2$ in
front of $\lambda$ comes from the factor $(1+B)$ in \eqref{31}.\\

In this way the Kitaev-Ising Hamiltonian turns into the rather
well-studied 2D Ising model in transverse field. The
ferromagnetic order is controlled by the Ising coupling which
tries to align all the virtual spins in the $+z$ or $-z$
direction. The transverse magnetic field controlled by $J$
competes with the Ising interaction and destroys the order if $J$
passes a critical value $J_c$. Density matrix renormalization
group \cite{ref13new} gives a value of the critical magnetic field
as $J_c\approx 6\lambda$. A simple mean field analysis (provided
in the appendix) gives the value $J_c=8\lambda$. When $J>J_c$,
the virtual spins try to align in the $+x$ direction. It is
important to note that in the limit $J=0$ (or $\lambda\lo
\infty$), the ground state of the virtual spin system is doubly
degenerate, while in the limit $J\lo \infty$ (or $\lambda=0$) the
ground state is unique and non-degenerate. What is interesting is
that these two phases of virtual spins correspond to the
topological and ferromagnetic phases of the actual spins on the
edges of the lattice.\\

To see this correspondence, consider one of the Hamiltonians, say
$H_A$. The $J$ term tends to align all the virtual spins in the
$+x$ direction, while the $\lambda$ term tends to align them in
the positive or negative $z$ direction. In the limit $\lambda\lo
0$ we have a unique ground state $|\tilde{+},\tilde{+},\cdots
\tilde{+}\ra$, while in the limit $\lambda\lo \infty$, there are
two ground states $|\tilde{0},\tilde{0},\cdots \tilde{0}\ra$ and
$|\tilde{1},\tilde{1},\cdots \tilde{1}\ra$. The same thing happens
in sublattice $B$. Therefore in the limit $\lambda\lo 0$, there
is a unique state, denoted by $(\tilde{+}_A,\tilde{+}_B)$, which
is the topological ground state of the Kitaev state
$|\Phi^{++}\ra$, while in the limit $\lambda\lo \infty$, there
are two ground states
$(\tilde{0}_A,\tilde{0}_B)=(\tilde{1}_A,\tilde{1}_B)$ and
$(\tilde{0}_A,\tilde{1}_B)=(\tilde{1}_A,\tilde{0}_B)$ which are
the ferromagnetically ordered states, where all virtual spins are
either up or down in the $z$ direction. Let us show this in a more
explicit way. Consider the state denoted by
$(\tilde{+}_A,\tilde{+}_B)$. In view of the notation \eqref{30},
and the fact that $|\tilde{+}\ra\propto
|\tilde{0}\ra+|\tilde{1}\ra$ the state of actual spins
corresponding to this state is given by $|\Phi^{++}\ra:=
  \prod_{i\in V}(1+A_i)|0\ra^{\otimes E},
$
which is a common eigenstate of all the $A_i$ and $B_i$ operators
with eigenvalue $1$, hence a ground state of the pure Kitaev
model. \\

In the other limit, the ground state $(\tilde{0}_A,\tilde{0}_B)$
denotes the state \eqref{30}, where none of the $A_i$'s act on
the state $|0\ra^{\otimes |E|}$, hence this is nothing but a
uniformly ordered ferromagnetic state in which all the spins are
up in the $z$ direction, i.e. $|0\ra^{\otimes |E|}$. We remind
the reader that
$(\tilde{0}_A,\tilde{0}_B)=(\tilde{1}_A,\tilde{1}_B)$, due to the
constraint $\prod_{i\in V}A_i=1$. In other words, if we act on
the state $|0\ra^{\otimes |E|}$ by all the vertex operators
$A_i$, nothing happens since the flipping actions of $A_i$
operators on the sublattice A are neutralized by those on
sublattice B. To flip all the spins, one needs to apply the
vertex operators $A_i$ on only one sublattice, hence the states
$(\tilde{0}_A,\tilde{1}_B)=(\tilde{1}_A,\tilde{0}_B)$ correspond
to the
ferromagnetically ordered state $|1\ra^{\otimes |E|}$.\\

\section{Topological characteristics; estimates of Wilson
loops}\label{wilson} In order to justify the transition from
topological to ferromagnetic order, we can estimate the value of
a Wilsonian loop,
\begin{equation}\label{Wc}
\la W_{C}\ra:=\la \prod_{i\in C}\sigma_{i,x}\ra,
\end{equation}
where the expectation value is calculated in the ground state and
$C$ is a closed curve on the dual lattice enclosing an area $S$,
i.e. $C=\partial S$. Let us denote the perimeter of $\partial S$
by $|\partial S|$ and the area of ${S}$ by $|S|$. Then it is
known that in the topological phase, the expectation value of
this Wilson loop behaves as $e^{\beta |\partial S|}$, while in
the non-topological phase it behaves as $e^{\gamma|S|}$, where
$\beta$ and $\gamma$ are two constants \cite{HammaWilson}. The
mapping of the Kitaev-Ising model to the 2D ITF model allows to
obtain estimates of the Wilson loop in the two regimes
perturbatively. To proceed we first note that in view of the
definition of the vertex operators $A_s$ in (\ref{2}), the
operator $W_{C}$ can be written as
\begin{equation}\label{WcX}
  W_{C}=\prod_{s\in S} A_s\equiv \prod_{s\in S}X_s,
\end{equation}
where in the last equality we have used the equivalence of $A_s$
with $X_s$ on virtual spins (see the paragraph after Eq.
\ref{30}), where the Hamitonian becomes a simple ITF Hamiltonian
as in (\ref{33}), we have to calculate the following expectation
\begin{equation}\label{WcPsi}
  \la  W_{C}\ra=\frac{\la \Psi|\prod_{s\in S} X_s|\Psi\ra}{\la \Psi|\Psi\ra}.
\end{equation}
where $|\Psi\ra$ is the ground state of the 2D ITF model with the
Hamiltonian given in (\ref{33}). Note that in view of the
decoupling between the two sublattices, we only consider one
sublattice. Consider now the two limits, near-Kitaev and
near-Ising separately. \\

\subsection{Close to the Kitaev limit}

In this limit, where $\gamma:=\frac{\lambda}{J}<< 1$, we can take
the Ising term in (\ref{33}) as a perturbation to the magnetic
field and approximate the ground state $|\Psi\ra$ as a  series

\begin{equation}\label{K1}
|\Psi\ra=|\Psi_0\ra+\gamma|\Psi_2\ra+\gamma^2(|\Psi'_2\ra+|\Psi_4\ra)+\cdots.
\end{equation}
Here $|\Psi_0\ra=|+\ra^{\otimes N}$ is the ground state in the
limit $\lambda=0$ and $|\Psi_2\ra$ denotes the linear
superposition of all states in which two adjacent spins have been
flipped by the $Z_iZ_j$ terms. Note since we are doing an
estimate and also we do not assume these states to be normalized,
all numerical factors coming from perturbation expansion like
energy differences and so on are absorbed in the definition of
these states. Similarly, $|\Psi_4\ra$ is the linear superposition
of all states in which $4$ (two pairs of nearest-neighbor) spins
have been flipped due to $(Z_iZ_j)(Z_kZ_l)$ terms and
$|\Psi'_2\ra$ is the linear superposition of all states in which
only two non-adjacent spins have been flipped by
$(Z_iZ_j)(Z_jZ_k)$ term and so on. We then have
\begin{equation}\label{K2}
\la \Psi |\Psi\ra=\la \Psi_0 |\Psi_0\ra+\gamma^2\la \Psi_2
|\Psi_2\ra+O(\gamma^3).
\end{equation}
We now note that $|\Psi_2\ra$ can be broken up into three kinds
of states, i.e.

\begin{equation}\label{K3}
  |\Psi_2\ra=|\Psi_2\ra_{_{\overline{S}}}+|\Psi_2\ra_{_{\underline{S}}}+|\Psi_2\ra_{_{\partial S}},
\end{equation}
where these states are described in figure (6). In view of this
figure and the fact that $X_i|\pm\ra=\pm |\pm\ra$, we then have
\begin{equation}\label{K4}
  W_C|\Psi_2\ra=|\Psi_2\ra_{_{\overline{S}}}+|\Psi_2\ra_{_{\underline{S}}}-|\Psi_2\ra_{_{\partial
  S}}.
\end{equation}
Combining (\ref{K3}) and (\ref{K4}), and keeping all the terms up
to order $\gamma^2$

\begin{equation}\label{K5}
   \frac{\la \Psi| W_{C}|\Psi\ra}{\la
   \Psi|\Psi\ra}=1-2\gamma^2\ \ _{_{\partial S}}\la \Psi_2|\Psi_2\ra_{_{\partial
   S}}+O(\gamma^3)
\end{equation}
and since $_{_{\partial S}}\la \Psi_2|\Psi_2\ra_{_{\partial
   S}}=|\partial S|$, we find that close to the Kitaev limit, we have
\begin{equation}\label{K6}
   {\frac{\la \Psi| W_{C}|\Psi\ra}{\la
   \Psi|\Psi\ra}} \approx e^{-2\gamma^2 |\partial
   S|}.
\end{equation}
Therefore as expected close to the Kitaev limit, the expectation
value of the Wilson loop behaves as the exponential of the
perimeter of the loop, which is characteristic of the topological
phase.

\begin{figure}[t]
\centering\vspace{1cm}
\includegraphics[width=12cm,height=4cm,angle=0]{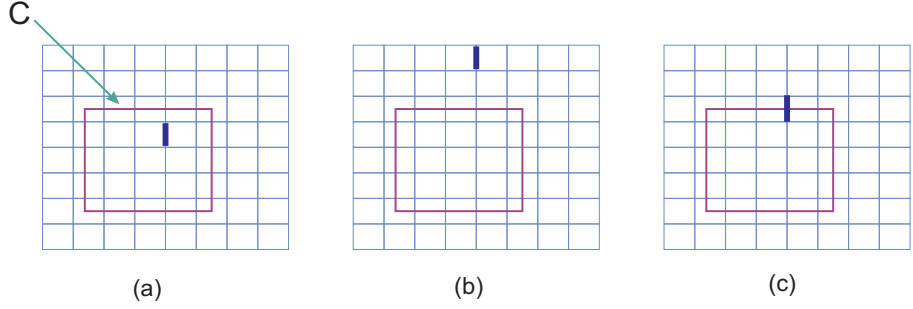}\vspace{1cm}
\caption{(Color Online) Different contributions to the state
$|\Psi_2\ra$. All the spins are in the state $|+\ra$, except the
two spins at the end of the bold link which have been flipped by
the $ZZ$ interaction and are in the state $|-\ra$.  Figures (a),
(b) and (c) exemplify contributions to the states
$|\Psi_2\ra_{\underline{S}}$, $|\Psi_2\ra_{\overline{S}}$ and
$|\Psi_2\ra_{\partial{S}}$, respectively.}
\end{figure}\label{PsiA}

\subsection{Close to the Ising limit}

We now consider the Ising limit where
$\frac{J}{\lambda}:=\frac{1}{\gamma}\leq 1$. Right at the Ising
point, consider one of the degenerate ground states, say one in
which all the spins are in the $+z$ direction, or using the
quantum computation terminology, all the spins are in the state
$0$. The magnetic field in (\ref{33}) perturbs this uniform ground
state by flipping spins one by one. These spins can be inside and
or outside the loop $\partial S$. Denote the lattice points inside
the loop by $S$ and the lattice points outside it by
$\overline{S}$. Let us $|\phi_k\ra|\chi_k\ra$ denote the product
state in which $|\phi_k\ra$ is the state pertaining to $S$, and
is the uniform superposition of all basis states in which exactly
$k$ spins have been flipped to $|1\ra$ (or $|-z\ra$) and
$|\chi_k\ra$ is the state pertaining to $\overline{S}$ in which
any number of spins have been flipped. Therefore the state
$|\phi_k\ra\chi_k\ra$ is a state in which at least $k$ spins have
been flipped. Then we can write the perturbative ground state as

\begin{equation}\label{Is1}
  |\Phi\ra=\sum_{k=0}^{|S|}\gamma^{-k}|\phi_k\ra|\chi_{k}\ra.
\end{equation}
Note that $|\chi_k\ra$ is a state which is normalized to $O(1)$.
The reason comes from perturbation theory, that is $ |\chi_k\ra$
is the superposition of states in $\overline{S}$ in which 0, 1,
or more spins have been flipped. Note also that
$$\la \phi_k|\phi_k\ra=\left(\begin{array}{c} |S|\\ k\end{array}\right). $$
Since the operator $W_C=\prod_{i\in S}X_i$ flips all the spins
inside $S$, we have
\begin{equation}\label{Is2}
  W_C|\Phi\ra=\sum_{k=0}^{|S|}\gamma^{-k}|\phi_{|S|-k}\ra|\chi_{k}\ra.
\end{equation}
Therefore we find
\begin{eqnarray}\label{Is3}
  \la \Phi|\Phi\ra &=&\sum_{k=0}^{|S|} \gamma^{-2k}\  \la \phi_k|\phi_k\ra \ \la
  \chi_k|\chi_k\ra\cr
&\approx& \sum_{k=0}^{|S|} \gamma^{-2k}\  \left(\begin{array}{c}
|S|\\ k\end{array}\right)\approx (1+\gamma^{-2})^{|S|}.
\end{eqnarray}
On the other hand we find from (\ref{Is2}) that
\begin{eqnarray}\label{Is4}
  \la \Phi|W_C|\Phi\ra &=&\sum_{k=0}^{|S|} \gamma^{-|S|}\ \la \phi_k|\phi_k\ra\  \la
  \chi_k|\chi_k\ra\cr
&\approx& \gamma^{-|S|}\sum_{k=0}^{|S|}  \left(\begin{array}{c}
|S|\\ k\end{array}\right)\approx \gamma^{-|S|}2^{|S|}.
\end{eqnarray}
Dividing (\ref{Is4}) by (\ref{Is3}) we find
\begin{equation}\label{Is5}
   {\frac{\la \Phi| W_{C}|\Phi\ra}{\la
   \Phi|\Phi\ra}} \approx \left(
   \frac{2\gamma^{-1}}{1+\gamma^{-2}}\right)^{|S|}=e^{-|S|\ln
   \frac{1+\gamma^2}{2\gamma}}.
\end{equation}
Therefore we have shown that close to the Kitaev and the Ising
points, the Wilson loop behaves as expected, that is,  its
logarithm is proportional to the perimeter of the loop in the
topological phase and proportional to the area in the
ferromagnetic phase. All this has been made possible by mapping
the system to the 2D Ising model in transverse field.

\section{Discussion}
We have introduced the Kitaev-Ising model, equation \eqref{13} as
a model for studying the transition between topological order and
ferromagnetic order in a lattice system. In particular we have
shown that on the quasi-one dimensional system of the ladder
(with periodic boundary condition), there is no quantum
transition between these two kinds of order at finite $\lambda$,
while in two dimensions a transition occurs at finite $\lambda$.
This is reminiscent of what we have for thermal phase transitions
based on symmetry breaking of discrete symmetries.\\

In the quasi-one dimensional case, we have exactly mapped the
problem to the problem of finding the ground state of an XY chain
in zero magnetic field for which exact solution by free fermion
techniques is available. On a two dimensional lattice on the
other hand, we have mapped the ground sector of the Kitaev-Ising
Hamiltonian to two copies of Ising models in transverse magnetic
fields, each defined on one sublattice, the latter model known to
show sharp transition for finite $\lambda$. Although we have not
attempted  a numerical study of the model near the transition
point, the equivalence with the 2D ITF model combined with the
analysis of the degenerate structure of the ground states and
their global properties in the two limits show that such a
transition does occur for some finite $\lambda$. We have also
estimated the Wilson loops and have shown that close to the
Kitaev and Ising points, the logarithm of the expectation value
of a Wilson loop is proportional to the perimeter of the loop in
the topological phase and to the area enclosed by the loop in the
ferromagnetic phase. It is also worth noticing an intriguing
difference between the characteristic of the two different
phases. In the topological phase, the four ground states are
distinguished by loop operators $T^1_z$ and $T_z^2$ and are
mapped to each other again by loop operators $T_x^1$ and $T_x^2$.
In the ferromagnetic phase on the other hand, the two degenerate
ground states are distinguished by a local operator $\sigma_{z}$,
while the two ground states are mapped to each other by a global
operator $\prod_{i\in E}\sigma_{i,x}$, encompassing the whole
lattice. Therefore during the transition, the distinguishing loop
operators shrink to points, while the transforming loop operators
expand to the whole
lattice.\\

This study can be extended in a few directions. First, one can
use numerical techniques to determine the ground state and its
properties as a function of the Ising coupling. Second it is
desirable to generalize the analysis of this paper to the cases
where the 2D lattice is not bipartite or the number of plaquttes
in the ladder is not even (the simplifying assumptions made here)
and to see if it leads to significantly different results.

\section{Acknowledgements:} We would like to thank
Saverio Pascazio, Razieh Mohseninia  and Luigi Amico for
interesting discussions during the early parts of this project.
V. K. thanks Abdus Salam ICTP for its associateship award and
support.

{}

\section{Appendix}
In this appendix we briefly do a mean field analysis of the 2D
Ising model in transverse field. Such an analysis reveals only a
very qualitative feature of the transition. Using a product trial
wave function $|\Psi\ra= |\phi\ra^{\otimes L}$ for the 2D ITF
model \eqref{33}, one needs to minimize the energy
$$\epsilon(|\phi\ra):=-J\la \phi|X|\phi\ra-4\lambda \la
\phi|Z|\phi\ra^2.$$ Taking $|\phi\ra=\cos
\frac{\theta}{2}|0\ra+\sin\frac{\theta}{2}e^{i\phi}|1\ra,$ leads
to the following expression
\begin{equation}\label{36}
  \epsilon(\theta, \phi)=-J\sin\theta \cos\phi - 4\lambda
  \cos^2\theta.
\end{equation}
Minimizing this energy, one obtains that the nature of the mean
field ground state changes at a critical value $\gamma_c=1$ (
$\gamma:=\frac{J}{8\lambda}$), that is the state which minimize
the mean field energy is

\begin{equation}\label{37}
  |\phi\ra=\left\lbrace\begin{array}{ll}
   \frac{1}{\sqrt{2}}(|0\ra+|1\ra), & \h \lambda <\frac{J}{8}\\ \cos \frac{\theta_\gamma}{2}|0\ra\pm \sin\frac{\theta_\gamma}{2}|1\ra & \h \frac{J}{8} < \lambda\end{array}\right.
\end{equation}
where
\begin{equation}\label{38}
  \sin \theta_\gamma=\frac{J}{8\lambda}.
\end{equation}
From this mean field analysis we find the following expectation
values $\la X\ra\equiv \la A_s\ra$ and $\la Z\ra\equiv \la
\sigma_z\ra$, also shown in figure (7).

\begin{equation}\label{39}
  \la X\ra\equiv \la A_s\ra=\left\lbrace\begin{array}{ll} 1,& \h \lambda < \frac{J}{8}\\ \frac{J}{8\lambda} &\h
   \frac{J}{8}< \lambda
   \end{array}\right.
\end{equation}
and

\begin{equation}\label{40}
  \la Z\ra\equiv \la \sigma_z\ra=\left\lbrace\begin{array}{ll} 0&\h \lambda < \frac{J}{8}\ \\ \pm \sqrt{ 1-(\frac{J}{8\lambda} )^2}& \h
   \frac{J}{8}< \lambda .\end{array}\right.
\end{equation}

\begin{figure}[t]
\centering\vspace{1cm}
\includegraphics[width=6cm,height=5.3cm,angle=0]{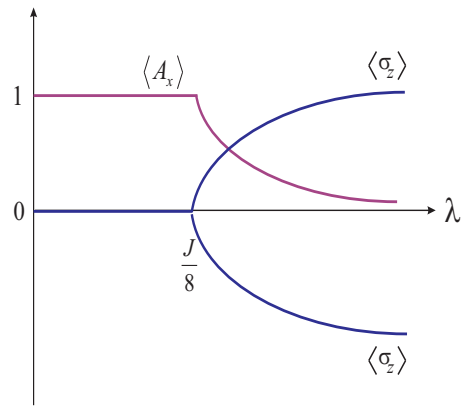}\vspace{1cm}
\caption{(Color Online) Mean field expectation values of $\la
X\ra$ and $\la Z\ra$ on the 2D Ising model in transverse field,
which are respectively equal to the $\la A_s\ra$ and $\la
\sigma_z\ra$ on the original lattice. Note that after the
transition point, the ground state becomes degenerate and the two
ground states can be distinguished by a local order parameter
$\la \sigma_z\ra$, hence two branches. The transition point
predicted by mean-field is $\frac{J}{8}$, perturbation theory
\cite{ref14} and renormalization method \cite{ref13new} give a
value $\approx \frac{J}{6}$.}
\end{figure}\label{fig6-XZ}

\end{document}